\documentclass[tradiabstract]{aa}

\usepackage{graphicx}
\usepackage{txfonts}

\usepackage{epsf}
\usepackage{rotating}
\usepackage{graphics}

\newcommand{\kepler}{\emph{Kepler}}
\def \MJ{M$_{\mathrm{Jup}}$}

\def \kms{km/s}
\def \ms{m\,s$^{-1}$}
\def \1s{$1\,\sigma$}

\def \t0{T$_0$}

\def\ms{\hbox{\,m\,s$^{-1}$}}        
  
\def\m2s2{\hbox{\,m$^{2}$\,s$^{-2}$}} 
\def\kms{\hbox{\,km\,s$^{-1}$}}     
\def\vsini{\hbox{$v$\,sin\,$i_*$}}  
  
\def\Msun{\hbox{$\mathrm{M}_{\odot}$}}       
\def\Rsun{\hbox{$\mathrm{R}_{\odot}$}}
\def\Mjup{\hbox{$\mathrm{M}_{\rm Jup}$}}
\def\Rjup{\hbox{$\mathrm{R}_{\rm Jup}$}}

\usepackage{amstext}
\usepackage{color}

\begin{document}

   \title{SOPHIE velocimetry of \textit{Kepler} transit candidates}
   \subtitle{XIX. The  transiting temperate giant planet KOI-3680b}

   \author{
G.~H\'ebrard\inst{1,2}
\and A.\,S.~Bonomo\inst{3}
\and R.\,F. D\'{\i}az\inst{4}
\and A.~Santerne\inst{5} 
\and N.\,C.~Santos\inst{6,7}
\and J.-M.~Almenara\inst{8}
\and S.\,C.\,C.~Barros\inst{6}
\and I.~Boisse\inst{5}
\and F.~Bouchy\inst{9}
\and G.~Bruno\inst{5,10}
\and B.~Courcol\inst{5}
\and M.~Deleuil\inst{5}
\and O.~Demangeon\inst{6}
\and T.~Guillot\inst{11}
\and G.~Montagnier\inst{1,2}
\and C.~Moutou\inst{12}
\and J.~Rey\inst{9}
\and P.\,A.~Wilson\inst{1,13,14}
}

  \offprints{G. H\'ebrard (hebrard@iap.fr)}

   \institute{
Institut d'Astrophysique de Paris, UMR7095 CNRS, Universit\'e Pierre \& Marie Curie, 
98bis boulevard Arago, 75014 Paris, France 
\and
Observatoire de Haute-Provence, CNRS, Universit\'e d'Aix-Marseille, 04870 Saint-Michel-l'Observatoire, France
\and
INAF, Osservatorio Astrofisico di Torino, via Osservatorio 20, 10025, Pino Torinese, Italy
\and
CONICET, Universidad de Buenos Aires, Instituto de Astronom\'{\i}a y F\'{\i}sica del Espacio (IAFE), 
Buenos Aires, Argentina
\and
Aix Marseille Universit\'e, CNRS, CNES, LAM (Laboratoire d'Astrophysique de Marseille), 13388 Marseille, France
\and
Instituto de Astrof{\'\i}sica e Ci\^encias do Espa\c{c}o, Universidade do Porto, CAUP, Rua das Estrelas, 4150-762 Porto, Portugal
\and
Departamento\,de\,F{\'\i}sica\,e\,Astronomia,\,Faculdade\,de\,Ci\^encias,\,Universidade\,do\,Porto,\,Rua\,Campo\,Alegre,\,4169-007\,Porto,\,Portugal
\and
Universit\'e Grenoble Alpes, CNRS, IPAG, 38000 Grenoble, France
\and
Observatoire de Gen\`eve,  Universit\'e de Gen\`eve, 51 Chemin des Maillettes, 1290 Sauverny, Switzerland
\and
INAF, Osservatorio Astrofisico di Catania, via S. Sofia, 78, 95123 Catania, Italy
\and
Universit\'e C\^ote d'Azur, OCA, Lagrange CNRS, 06304 Nice, France
\and
Canada France Hawaii Telescope Corporation, Kamuela, HI 96743, USA
\and
Department of Physics, University of Warwick, Gibbet Hill Road, Coventry, CV4 7AL, UK
\and
Centre for Exoplanets and Habitability, University of Warwick, Gibbet Hill Road, Coventry CV4 7AL, UK
}

   \date{Received TBC; accepted TBC}
      
\abstract{
Whereas thousands of transiting giant exoplanets are known today, only a few are well characterized 
with long orbital periods. Here we present KOI-3680b, a new planet in this category. 
First identified by the \kepler\ team as a promising candidate from the photometry 
of the \kepler\ spacecraft, we establish here its planetary nature from the radial velocity follow-up 
secured over two years with the SOPHIE spectrograph at Observatoire de Haute-Provence, France.
The combined analysis of the whole dataset allows us to fully characterize this new planetary system. 
KOI-3680b has an orbital period of $141.2417 \pm 0.0001$~days, a mass of 
$1.93 \pm 0.20$\,\Mjup,\ and a radius of $0.99 \pm 0.07$\,\Rjup.
It exhibits a highly eccentric orbit ($e=0.50 \pm 0.03$) around an early G~dwarf. 
KOI-3680b is the transiting giant planet with the longest period characterized so far around 
a single star; it offers opportunities to extend studies which were mainly devoted to exoplanets 
close to their host stars, and to compare both exoplanet populations.
}

\keywords{Planetary systems -- Techniques: radial velocities -- Techniques: photometric -- 
   Techniques: spectroscopic -- Stars: individual: KOI-3680 (Kepler-1657)}

\authorrunning{H\'ebrard et al.}
\titlerunning{KOI-3680b}

\maketitle

\section{Introduction}
\label{sect_intro}

By continuously monitoring the light curves of 156\,000 stars from May 2009 to May 2013
with high photometric accuracy, the \kepler\ spacecraft revealed more than 4500 transiting planet 
candidates. These Kepler objects of interest (KOIs) have corresponding  
radii ranging from giant planets even larger than Jupiter down to Earth-like planets smaller 
than our Earth. They also cover a wide span of distances from their host stars, with 
orbital periods of several hours, days, or months.

Whereas several purely stellar configurations can mimic planetary transits, in particular 
those involving blended binaries, three main methods have been used to distinguish genuine 
planets from false positives in the KOI sample.
First, from Bayesian statistics of the \kepler\ light curves and comparison of astrophysical
configuration probabilities, it is possible to validate the planetary nature of a transit. 
About half of the KOIs were validated this way, including more than 800 transiting 
planets in multiple systems by Rowe et al.~(\cite{rowe14}; 
see also Lissauer et al.~\cite{lissauer12}, \cite{lissauer14}) 
and nearly 1300 additional planets by Morton et al.~(\cite{morton16}). 
That validation technique however does not allow the mass of the planets to be measured, 
which is a particularly important parameter for their characterization and study.

The second technique relies on measurements of transit timing variations (TTVs) in 
multi-planet systems and is also mainly based on the the \kepler\ light curves 
(e.g. Holman et al.~\cite{holman10}, Fabrycky et al.~\cite{fabrycky12}, 
Ford et al.~\cite{ford12}). Tens of KOIs were identified as planets and their masses measured 
using such dynamic analyses of mutual gravitational interactions.

The third method requires additional ground-based high-resolution spectroscopic 
observations. These allow the planetary nature of a KOI to be established or rejected, as well 
as the mass of the identified planets to be measured 
thanks to the induced stellar radial velocity (RV) variations.
This is the historical method to characterize transiting planet candidates 
from ground- and space-based photometric surveys, and was used to characterize
tens of KOIs (e.g. Bonomo et al.~\cite{bonomo14}, H\'ebrard et al.~\cite{hebrard14}, 
Marcy et al.~\cite{marcy14}, Buchhave et al.~\cite{buchhave16}).
Today, the exoplanet light curve in addition to TTV and RV analyses provides the most 
comprehensive way of obtaining exoplanetary properties, that is, measured
orbital and physical parameters of planets, including period, eccentricity, mass, and~radius.

An RV follow-up of KOIs was begun in 2010 with the SOPHIE spectrograph 
at the 193-cm Telescope of the Observatoire de Haute-Provence, France. 
The program allowed us to announce and characterize new transiting companions, including 
hot Jupiters 
(e.g. Santerne et al.~\cite{santerne11}, Bonomo et al.~\cite{bonomo12a}, Deleuil et al.~\cite{deleuil14}, 
Almenara et al.~\cite{almenara15}), 
brown dwarfs 
(e.g. D\'{\i}az et al.~\cite{diaz13}, Moutou et al.~\cite{moutou13}), 
and low-mass stars
(e.g. Ehrenreich et al.~\cite{ehrenreich11}, Bouchy et al.~\cite{bouchy11}, D\'{\i}az et al.~\cite{diaz14}), 
as well as to put constraints on systems presenting TTVs 
(Bruno et al.~\cite{bruno15}, Almenara et al.~\cite{almenara18}).
The occurrence rate and physical properties of giant planets with up to 400-day 
orbital periods were moreover constrained from this program, which also revealed a high false-positive 
rate among the corresponding KOIs (Santerne et al.~\cite{santerne12},~\cite{santerne16}).

The same program also allowed us to announce and characterize the new transiting giant planet 
KOI-1257b on a particularly long orbital period of 86.6~days (Santerne et al.~\cite{santerne14}). 
Only a few transiting planets are well characterized with periods of a few tens of days or longer.
Indeed, the geometric probability 
for the orbital plane to be aligned with the line of sight roughly scales with the power ${-5/3}$ of the 
orbital period (Beatty \&\ Gaudi~\cite{beatty08}).
For orbital periods around 100 days, the geometric transit probability 
is of the order of 1\,\%, and when the alignment is good enough to produce 
transits, opportunities to observe them are rare due to the long period. Thus, in addition 
to KOI-1257b, only a few transiting, long-period planets are well characterized today; 
examples include
HD\,80606b (Moutou et al.~\cite{moutou09}, H\'ebrard et al.~\cite{hebrard10}), 
CoRoT-9b (Deeg et al.~\cite{deeg10}, Bonomo et al.~\cite{bonomo17a}), or
Kepler-432b (Ciceri et al.~\cite{ciceri15})
characterized with long-time-span RVs, 
and the multiple systems 
Kepler-30 (Sanchis-Ojeda et al.~\cite{sanchis12}), 
Kepler-79 (Rowe et al.~\cite{rowe14}), and
Kepler-87 (Ofir et al.~\cite{ofir14}) 
characterized with TTVs.
Also classified using TTVs are some long-period, transiting planets orbiting binary stars, 
as in the systems
Kepler-16 (Doyle et al.~\cite{doyle11}), 
Kepler-34, and
Kepler-35 (Welsh et al.~\cite{welsh12}).

\begin{table}[b]
\centering
\caption{IDs, coordinates, and magnitudes of the planet-host star.}            
\begin{tabular}{lc}       
\hline 
Kepler Object of Interest & KOI-3680 \\
\kepler\ exoplanet catalog & Kepler-1657 \\
Kepler ID               & 9025971                       \\
2MASS ID        & J19330757+4518348             \\
Gaia ID         & 2126518783059463168           \\
\hline            
RA (J2000)   & 19:33:07.574  \\   
DEC (J2000) &  +45:18:34.81  \\  
\hline
Gaia parallax (mas)   &  $1.056 \pm 0.027$ \\ 
Distance (pc)   &  $950 \pm 25$ \\ 
RA PM (mas/yr)  &       $-11.51 \pm 0.04$ \\
DEC PM (mas/yr) &       $-3.60 \pm 0.04$ \\
\hline
Kepler magn. $K_{\rm p}$        & 14.524  \\
Howell Everett Survey Johnson-B & 15.47 \\
Howell Everett Survey Johnson-V & 14.77 \\
Gaia-$G$                &       $ 14.50269 \pm 0.00026$  \\
Gaia-$BP$       &       $ 14.880 \pm 0.002$  \\
Gaia-$RP$       &       $ 13.974 \pm 0.001$  \\
2MASS-$J$       &       $13.384 \pm 0.021$       \\ 
2MASS-$H$       &       $13.041 \pm 0.027$       \\
2MASS-$K_s$     &       $13.101 \pm 0.039$       \\
\hline
\end{tabular}
\label{startable_KOI}      
\end{table}

The vast majority of well-characterized transiting planets have short orbital periods and 
receive high irradiation from their host star. They are particularly interesting 
(e.g. Winn~\cite{winn10}) because  their physics as well as their formation 
and evolution processes can be constrained.
Indeed, it is possible to measure the bulk density of transiting planets, permitting 
studies of their internal structure, as well dynamic analyses through obliquity measurements, 
or through TTVs in cases of multiple systems. They also allow atmospheric studies 
in absorption through transits and in emission through occultation
for the brightest host stars.
Therefore, there is great interest in extending such powerful analyses from close-in planets 
to less irradiated ones. To do that, new transiting, long-period planets need to be detected and 
characterized. Long-period, giant planets present additional interest as some of them are the probable 
precursors of the observed and well-studied population of hot Jupiters. Studying more-temperate 
planets could also put important constraints on the migration processes and their effects on the 
internal structure and atmosphere of migrating planets. 

\begin{figure}[b]
\centering
\includegraphics[width=6.6cm, angle=90]{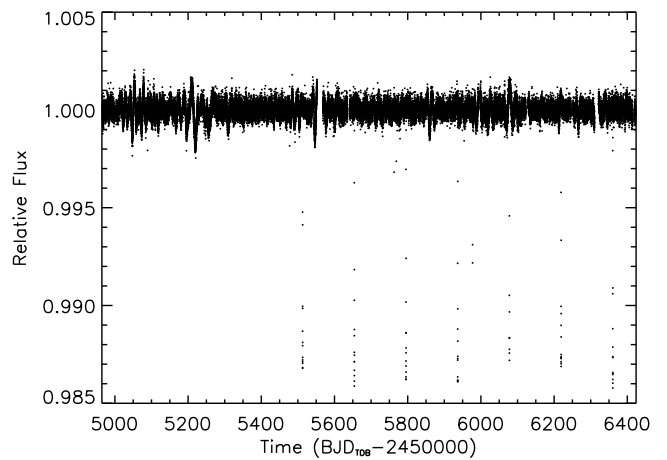}
\caption{\emph{Kepler} light curve showing the transits of KOI-3680.01 and a low-amplitude variability
 due to the rotational modulation of photospheric active regions. The first three transits unluckily 
 fell within gaps in the time series.}
\label{fig_LC}
\end{figure}

Here we present \object{KOI-3680b}, a new transiting, long-period planet. 
The planet candidate was revealed with an orbital period of 141~days 
by Wang et al.~(\cite{wang13}) and Rowe et al.~(\cite{rowe15}) as KOI-3680.01.
Its transit depth was characteristic of a giant planet.
Both analyses used the Quarters Q1 to Q12 representing three years of \kepler\ observations.
Furthermore, this candidate was not detected in the previous releases of \kepler\ data 
(Borucki et al.~\cite{borucki11a}, \cite{borucki11b}, Batalha et al.~\cite{batalha12}), 
illustrating the need for long time-span data to detect such long-period 
events. Here, an additional cause  was the fact that the first three transits unfortunately fell within gaps
in the time series (see Sect.~\ref{sect_kepler_photometry} and Fig.~\ref{fig_LC}).
Morton et al.~(\cite{morton16}) computed the false-positive probability of KOI-3680.01 to be 
$1.20 \pm 0.15$\,\%\ so could not reach a definitive conclusion about its~nature.
Therefore it remained undetermined whether these transits were 
caused by a planet or another scenario.

Our RV follow-up with SOPHIE allows us to show they are caused by a planet and to characterize
the parameters of the planetary system.
We describe the photometric and spectroscopic observations of the object in 
Sect.~\ref{sect_observations} and the analysis of the whole dataset and the results 
in Sect.~\ref{sect_analysis}. We discuss our results in Sect.~\ref{sect_discussion}
and conclude in~Sect.~\ref{sect_conclusion}.

\section{Observations and data reduction}
\label{sect_observations}
 
\subsection{Photometric detection with Kepler}
\label{sect_kepler_photometry}

The IDs, coordinates, and magnitudes of the star KOI-3680 are reported in Table~\ref{startable_KOI}. 
\emph{Kepler} photometric data were retrieved 
from the MAST archive\footnote{http://archive.stsci.edu/kepler/data search/search.php}; 
they were acquired only in long-cadence mode, 
that is, with a temporal sampling of 29.4 min over four years 
from 13 May 2009 to 11 May 2013 (quarters Q1 to Q17). 
We used the \emph{Kepler} light curves as reduced by both the Simple Aperture Photometry (SAP)
and Pre-search Data Conditioning (PDC) \emph{Kepler} pipelines (Jenkins et al.~\cite{jenkins10}).
Those data are described in the DR25 \kepler\ catalogue (Thompson et al.~\cite{thompson18}).

The PDC light curve, which is shown in Fig.~\ref{fig_LC}, 
clearly presents seven transits with a depth of about 1.4\,\%. 
Three more transits should have been observed by \emph{Kepler} 
at $\rm BJD_{\rm TDB} -2\,450\,000 < 5500$ (see Fig.~\ref{fig_LC}) 
but, unfortunately, they all fell in data gaps caused by \emph{Kepler} operational activities, 
such as safe modes and pointing tweaks (Jenkins et al.~\cite{jenkins10}). 
Moreover, the fifth transit observed at 6077.8960 $\rm BJD_{\rm TDB}-2\,450\,000$ 
misses the egress and is thus partial.  
We note that individual transits seem to vary in depth in Fig.~\ref{fig_LC}. This is 
mainly due to transits occurring during small bumps or ditches in the light curve due to 
stellar activity not yet corrected at that stage. 
To a less extent this could also be due to possible residual background as well as 
possible instrumental effects. In the analysis presented below in 
Sect.~\ref{sect_Parameters_planetary_system}, the phase-folded curve 
is normalized around each transit to correct for stellar activity. 
The final uncertainty on the planetary radius 
accounts for possible residual, though not significant, differences in the depth 
of individual transits. 

After removing the transits of KOI-3680.01, we searched for additional transit signals due to 
possible coplanar planetary companions using a modified version of the pipeline described by
Bonomo et al.~(\cite{bonomo12b}) but found none; 
there is thus no evidence for a multiple transit~system.

\begin{figure}[h]
\centering
\includegraphics[width=6.6cm, angle=90]{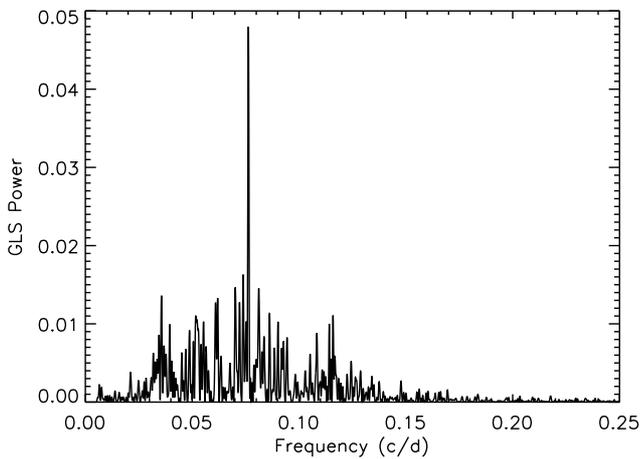}
\caption{Generalized Lomb-Scargle periodogram of the \emph{Kepler} light curve. 
The highest peak at $\nu=0.0763$~c/d, which corresponds to a period of 13.10~d, 
is likely related to the stellar rotation period.}
\label{fig_GLS_LC}
\end{figure}

The \emph{Kepler} light curve also shows low-amplitude variability with a maximum peak-to-peak 
variation of $\sim 0.4\,\%$ (Fig.~\ref{fig_LC}), which is likely due to the rotational modulation of 
photospheric starspots and/or faculae. 
Despite the low amplitude, the Generalized Lomb-Scargle (GLS) periodogram (Zechmeister \&\ 
Kurster~\cite{zechmeister09})
of the light curve uncovers a significant periodicity at $13.102 \pm 0.002$~d (Fig.~\ref{fig_GLS_LC}). 
This periodicity could be related to the stellar rotation period 
(see below Sect.~\ref{sect_stellar_parameters}).
Following the approach of Czesla et al.~(\cite{czesla09}), we estimate the 
impact of unocculted starspots on the transit depth to be
24\,ppm at most, which is negligible here.

After removing the transits and low-frequency stellar variations, the r.m.s. of the light curve
is 245\,ppm, which is almost equal to the average of the uncertainties of the photometric 
measurements, that is, 235\,ppm.
So there are no signs of remaining noise or additional signals.

\begin{figure*}[h!]
\centering
\begin{minipage}{16cm}
\includegraphics[width=6.3cm, angle=90]{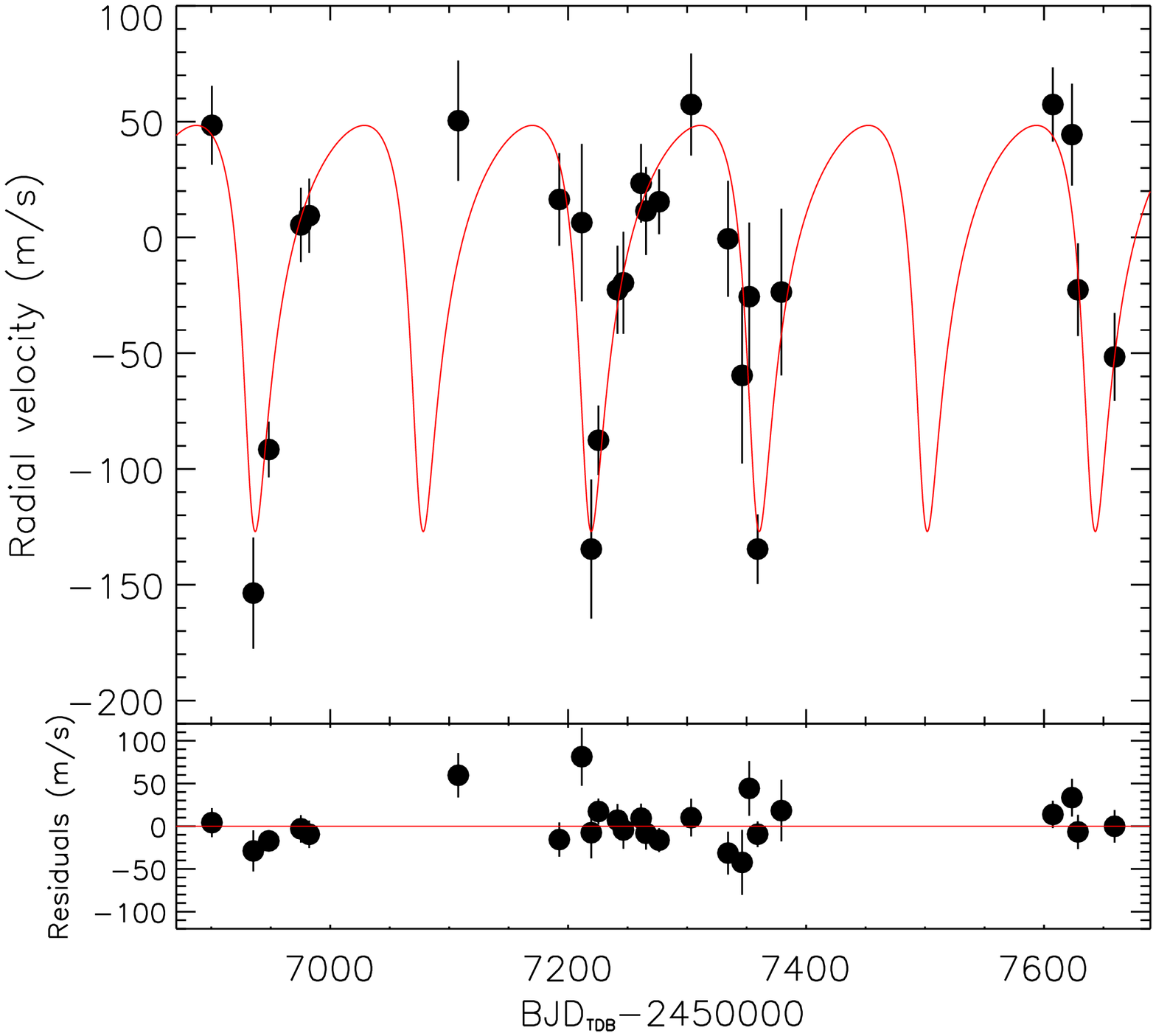}
\includegraphics[width=6.3cm, angle=90]{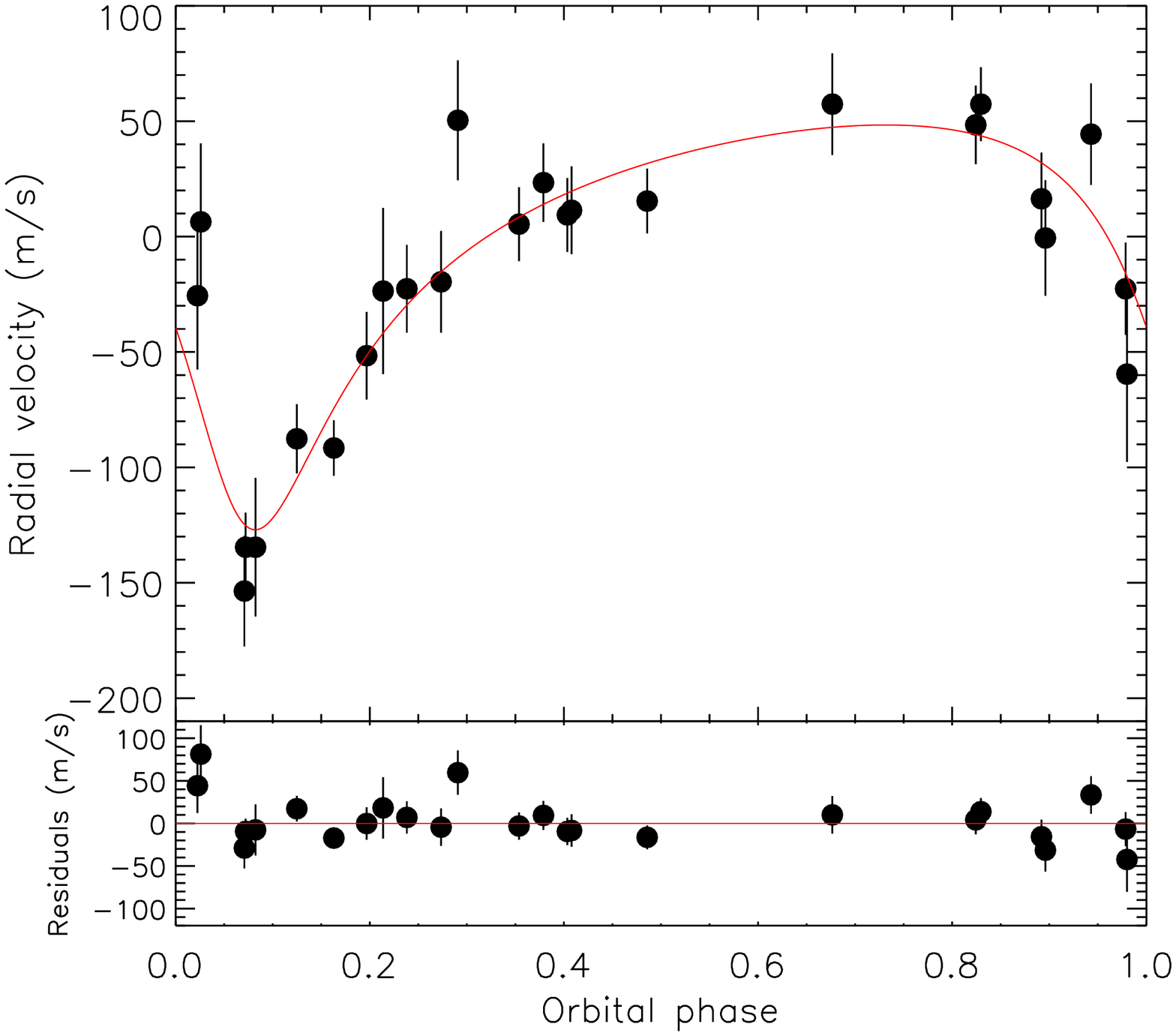}
\vspace{0.3cm}
\caption{
\emph{Left panel}: SOPHIE radial velocities of KOI-3680 as a function of time 
and the best Keplerian model (red solid line).
\emph{Right panel}: As in the left panel but as a function of the orbital phase 
(transits occur at phase equal to zero/one). On both plots the lower panel displays the residuals.}
\label{fig_RV}
\end{minipage}
\end{figure*}

\subsection{Radial-velocity follow-up with SOPHIE}

We observed KOI-3680 with the SOPHIE spectrograph, first
to establish the planetary nature of the transiting candidate, then to characterize the 
secured planet by measuring in particular its mass and orbital eccentricity. 
SOPHIE is dedicated to high-precision 
RV  measurements at the 1.93-m telescope of the Haute-Provence Observatory
(Perruchot et al.~\cite{perruchot08}, Bouchy et al.~\cite{bouchy09a},~\cite{bouchy13}).
We used its High-Efficiency mode with a resolving power 
$R=40\,000$ and slow readout mode to increase the throughput for this faint star.
We obtained 28 observations between August 2014 and September 2016
but three of them were not used due to their poor quality.
Exposure times were 60 minutes, except for five observations where it was slightly shorter.
The signal-to-noise ratios (S/N)\ per pixel at 550~nm range between 14 and 25 depending on
the exposure and weather conditions (Table~\ref{table_rv}).

\begin{table}[h]
\caption{SOPHIE measurements of the planet-host star KOI-3680}
\begin{center}
\begin{tabular}{cccrrr}
\hline
BJD$_{\rm UTC}$ & RV & $\pm$$1\,\sigma$ & bisect.$^\ast$ & exp. & SNR$^\dagger$ \\
-2\,450\,000 & (km/s) & (km/s) & (km/s)  & (sec) &  \\
\hline
6900.5292  &  -44.880  &  0.017  &  -0.069  &  3600  &  16.2 \\
6935.3558  &  -45.082  &  0.024  &   0.002  &  3600  &  16.3 \\
6948.3726  &  -45.020  &  0.012  &  -0.044  &  3600  &  23.2 \\
6975.3101  &  -44.923  &  0.016  &  -0.018  &  3600  &  20.0 \\
6982.3470  &  -44.919  &  0.016  &  -0.054  &  3600  &  21.9 \\
7107.6489  &  -44.878  &  0.026  &   0.002  &  3164  &  18.2 \\
7192.5683  &  -44.912  &  0.020  &   0.005  &  3600  &  16.7 \\
7211.4976  &  -44.922  &  0.034  &  -0.027  &  3600  &  11.9 \\
7219.4596  &  -45.063  &  0.030  &  -0.088  &  2452  &  12.4 \\
7225.4649  &  -45.016  &  0.015  &  -0.017  &  3600  &  19.8 \\
7241.4781  &  -44.951  &  0.019  &  -0.027  &  3600  &  18.1 \\
7246.4579  &  -44.948  &  0.022  &  -0.051  &  3600  &  16.1 \\
7261.3529  &  -44.905  &  0.017  &  -0.047  &  3600  &  18.3 \\
7265.4376  &  -44.917  &  0.019  &   0.026  &  3600  &  18.1 \\
7276.4394  &  -44.913  &  0.014  &   0.024  &  3600  &  17.3 \\
7303.3809  &  -44.871  &  0.022  &   0.050  &  3600  &  13.5 \\
7334.3829  &  -44.929  &  0.025  &  -0.034  &  3443  &  17.4 \\
7346.2494  &  -44.988  &  0.038  &   0.039  &  2704  &  10.8 \\
7352.2475  &  -44.954  &  0.032  &  -0.046  &  3600  &  14.1 \\
7359.2618  &  -45.063  &  0.015  &  -0.032  &  3600  &  19.1 \\
7379.2595  &  -44.952  &  0.036  &  -0.002  &  2801  &  15.0 \\
7607.4763  &  -44.871  &  0.016  &  -0.054  &  3600  &  15.8 \\
7623.5069  &  -44.884  &  0.022  &  -0.029  &  3600  &  16.0 \\
7628.5428  &  -44.951  &  0.020  &  -0.030  &  3600  &  14.4 \\
7659.3588  &  -44.980  &  0.019  &  -0.085  &  3600  &  14.3 \\
\hline
\multicolumn{6}{l}{$\ast$: bisector spans; error bars are twice those of the RVs.} \\ 
\multicolumn{6}{l}{$^\dagger$: signal-to-noise ratio per pixel at 550\,nm.} \\
\end{tabular}
\end{center}
\label{table_rv}
\end{table}

The spectra were extracted using the SOPHIE pipeline (Bouchy et al.~\cite{bouchy09a}) 
and the radial velocities were measured from the weighted cross-correlation with a 
numerical mask characteristic of the spectral type of the star (Baranne et al.~\cite{baranne96}, 
Pepe et al.~\cite{pepe02}). 
They were corrected from the CCD charge transfer inefficiency (Bouchy et al.~\cite{bouchy09b})
and their error bars were computed from the cross-correlation 
function (CCF) using the method presented by Boisse et al.~(\cite{boisse10}). 
Following the procedure adopted by Santerne et al~(\cite{santerne16}), we 
used a G2-type mask and corrected for instrumental drifts in the RV using those 
measured on the constant star HD\,185144 on the same~nights.
The causes for these instrumental drifts are not well understood or
identified but might be due 
to thermal effects. Their dispersion is 5\,m/s over the nights where KOI-3680 was 
observed; we corrected for this effect despite its low amplitude~because that 
correction does not add significant extra RV~uncertainty.

\begin{figure}[h] 
\begin{center}
\includegraphics[scale=0.7]{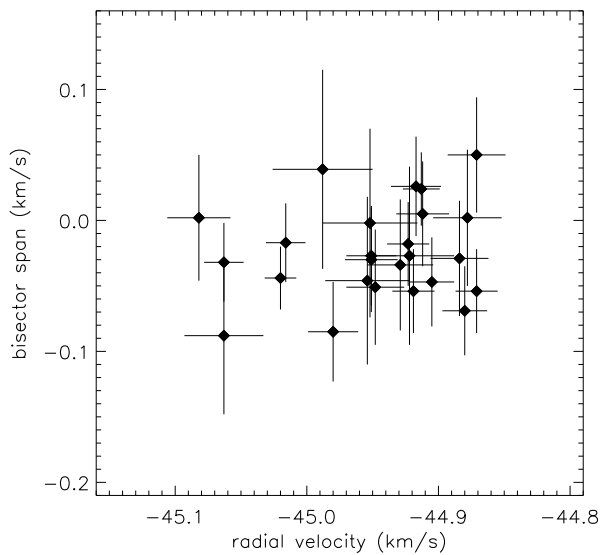}
\caption{Bisector span as a function of the radial velocities with 1-$\sigma$\,Êerror bars. 
The ranges extend equally on the $x$ and $y$ axes.
The bisector spans show a smaller dispersion than the RVs, and no correlation is seen
between the two quantities.}
\label{fig_bis}
\end{center}
\end{figure}

Nine spectra were polluted by moonlight but the RV of the Moon signal 
was shifted by at least 32\,km/s from the RV of KOI-3680.  Such 
shifts are large enough to avoid significant perturbation on the stellar RV.
Indeed, the median moonlight correction 
(done thanks to the second SOPHIE aperture located on the sky 
background 2' away from the first aperture located on the star; see
e.g. H\'ebrard et al.~\cite{hebrard08}, Bonomo et al.~\cite{bonomo10})
here is 15\,m/s which is smaller than the 
typical accuracy of our RV measurements. As the moonlight correction could
add extra uncertainty on the final RV
(due to the noise present in the second-aperture spectra), we decided not to correct for this
pollution. As an additional test, we checked that the Keplerian solution ignoring 
these nine Moon-polluted spectra agrees within 1\,$\sigma$ with our final 
solution using the whole dataset.

The resulting CCFs have full width at half maximum (FWHM) of $9.9 \pm 0.1$\,km/s 
and contrast that represents $\sim30$\,\%\ of the continuum. 
The lines are slightly broader than what is usually measured in High-Efficiency mode 
due to the stellar rotation of KOI-3680 (we measure \vsini$=3.5 \pm 1.0 $\,\kms\
from  the CCF width; see Sect.~\ref{sect_stellar_parameters} below).
The radial velocities have accuracies ranging
from $\pm12$ to $\pm38$\,m/s depending on the S/N, with 
a median value of $\pm20$\,m/s. 
They are reported in Table~\ref{table_rv} and displayed in Fig.~\ref{fig_RV} together with their 
Keplerian fits and their residuals. They show significant variations in phase with the \kepler\
transit ephemeris, for an eccentric orbit and a semi-amplitudes of the order of 
100~\ms. This implies a companion mass in the planet regime.

Radial velocities measured using different stellar masks (F0, K0, or K5) produce 
variations with similar amplitudes 
(in agreement within 0.8, 0.7, and 1.1\,$\sigma$, respectively)
to those obtained with the G2 mask, so it is unlikely 
that these variations are produced by blend scenarios 
composed of stars of different spectral types.
Similarly, the measured CCF bisector spans (Table~\ref{table_rv})  
quantify  possible shape variations of the spectral lines. They show 
a dispersion of 36\,m/s which is 1.7 times smaller than the RV dispersion, 
whereas each bisector span is roughly two times less precise than the 
corresponding RV measurement. Therefore, the bisector spans show no significant variations.
Moreover, they show no correlations with the RVs as shown in Fig.~\ref{fig_bis}.
The linear correlation parameter is $+0.10 \pm 0.12$ and 
the Pearson and Spearman's rank correlation factors 
have low values of 0.18 and 0.12, respectively.
This reinforces the conclusion that the RV variations are not caused by 
spectral-line profile changes attributable to blends or stellar activity.
We therefore conclude that the event KOI-3680.01 is due to a transiting planet, 
which we designate as KOI-3680b~hereafter.

\section{System characterization}
\label{sect_analysis}

\subsection{Spectral analysis of the host star}
\label{sect_stellar_parameters}

Stellar atmospheric parameters ($T_{\rm eff}$, $\log{g}$, microturbulence $\xi_{\rm t}$ and [Fe/H]) 
and respective error bars were derived using the methodology described in 
Sousa et al.~(\cite{sousa08}) and Santos et al.~(\cite{santos13}).
In brief, we make use of the 
equivalent widths of 153 \ion{Fe}{I} and 22 \ion{Fe}{II} lines, and we assume ionization and 
excitation equilibrium. The process makes use of a grid of Kurucz model 
atmospheres (Kurucz~\cite{kurucz93}) and the radiative transfer code MOOG 
(Sneden~\cite{sneden73}). 

The equivalent widths were measured on a SOPHIE spectrum built from the addition of the 
existing high-resolution spectra used for the RV measurements, but excluding the
SOPHIE spectra presenting Moonlight contamination. The resulting average SOPHIE spectrum 
has a S/N of $\sim$40 as measured in continuum regions near 6600A. We obtained 
$T_{\rm{eff}} = 5830 \pm 100$\,K, 
log\,$g = 4.49 \pm 0.13$ (cgs), 
 $\xi_{\rm t} = 1.20 \pm 0.13$\,\kms, and
 [Fe/H] = $0.16 \pm 0.07$.
Using the calibration of Torres et al.~(\cite{torres10}) with a correction following Santos 
et al.~(\cite{santos13}), we derive a mass and radius of 
$1.04 \pm 0.05$\,\Msun\ and $0.96 \pm 0.16$\,\Rsun, respectively.
Those values of $T_{\rm{eff}}$ and [Fe/H]
are used as input in the analysis below (Sect.~\ref{sect_Parameters_planetary_system}).
Our derived stellar parameters agree within 1\,$\sigma$ with those reported by the  
\kepler\ mission in the DR25 catalogue (Thompson et al.~\cite{thompson18}). 
DR25 values are
$T_{\rm{eff}} = 5705^{+104}_{-115}$\,K,         
log\,$g = 4.397^{+0.095}_{-0.105}$ (cgs),        [Fe/H] = $0.00 \pm 0.15$,              
$M_* = 0.941^{+0.068}_{-0.056}$\,\Msun, and $R_* = 1.017^{+0.147}_{-0.107}$\,\Rsun. 

We also derived the projected rotational velocity \vsini$=3.5 \pm 1.0 $\,\kms\ 
from the parameters of the CCF using the calibration of Boisse et al.~(\cite{boisse10}).
This implies a stellar rotation period upper limit $P_{\rm rot} < 13.9^{+5.5}_{-3.1}$~d, 
in agreement with the value $P_{\rm rot} \simeq 13.10$\,d derived above 
in Sect.~\ref{sect_kepler_photometry}.

Therefore, KOI-3680 is an early G dwarf.

\begin{table*}[]
\centering
\caption{KOI-3680 system parameters.}            
\begin{minipage}[t]{13.0cm} 
\renewcommand{\footnoterule}{}                          
\begin{tabular}{l l }        
\hline\hline                 
\emph{Stellar parameters}  &  \\
\hline
Star mass [\Msun] &  $ 1.01_{-0.16}^{+0.07} $  \\
Star radius [\Rsun] & $ 0.96_{-0.06}^{+0.05}$   \\
Stellar density $\rho_{*}$ [$ \rm g\;cm^{-3}$] & $1.88_{-0.34}^{+0.40}$ \\
Age $t$ [Gyr]  & $3.2_{-2.3}^{+9.6}$  \\
Effective temperature $T_{\rm{eff}}$[K] & 5830 $\pm$ 100 \\
Spectroscopic surface gravity log\,$g$ [cgs] &  $4.49 \pm 0.13$  \\
Derived surface gravity log\,$g$ [cgs] &  $4.473_{-0.034}^{+0.027}$ \\
Metallicity $[\rm{Fe/H}]$ [dex] & 0.16  $\pm$ 0.07 \\
Projected rotational velocity \vsini\ [\kms]  &   $3.5 \pm 1.0 $  \\
Kepler limb-darkening coefficient $q_{1}$  &  $0.30_{-0.07}^{+0.08}$ \\
Kepler limb-darkening coefficient $q_{2}$  &  $0.46^{+0.11}_{-0.08}$  \\
Kepler limb-darkening coefficient $u_{a}$  &  $0.51 \pm 0.05$ \\
Kepler limb-darkening coefficient $u_{b}$  &  $0.04 \pm 0.11$  \\
Systemic velocity  $V_{\rm r}$ [\kms] & $-44.9284 \pm 0.0044$ \\
\hline
\emph{Transit and orbital parameters}  &  \\
\hline
Orbital period $P$ [d] & $141.241671 \pm 0.000086$ \\
Transit epoch $T_{ \rm 0} [\rm BJD_{TDB}-2\,450\,000$]~$^a$ & $5936.65434 \pm 0.00018$  \\
Transit duration $T_{\rm 14}$ [d] & $0.2807 \pm 0.0012$  \\
Radius ratio $R_{\rm p}/R_{*}$ & $0.10646_{-0.00059}^{+0.00079}$   \\
Inclination $i_p$ [deg] & $89.892_{-0.086}^{+0.069}$  \\
$a/R_{*}$ & $125.7 \pm 8.4$  \\
Impact parameter $b$ & $0.15^{+0.10}_{-0.09}$  \\
$\sqrt{e}~\cos{\omega}$ &  $0.300_{-0.091}^{+0.086} $ \\
$\sqrt{e}~\sin{\omega}$  &  $-0.635_{-0.037}^{+0.046} $ \\
Orbital eccentricity $e$  &  $0.496 \pm 0.031$   \\
Argument of periastron $\omega$ [deg] & $154.7 \pm 7.8$ \\
Radial-velocity semi-amplitude $K$ [\ms] & $87.7 \pm 6.6$ \\
\hline
\multicolumn{2}{l}{\emph{Planetary parameters}} \\
\hline
Planet mass $M_{\rm p} ~[\Mjup]$  &  $1.93_{-0.21}^{+0.19}$  \\
Planet radius $R_{\rm p} ~[\Rjup]$  &  $0.99^{+0.06}_{-0.07}$  \\
Planet density $\rho_{\rm p}$ [$\rm g\;cm^{-3}$] &  $2.46^{+0.42}_{-0.36}$  \\
Planet surface gravity log\,$g_{\rm p }$ [cgs] &  $3.69 \pm 0.05$  \\
Orbital semi-major axis $a$ [AU] & $0.534_{-0.030}^{+0.012}$   \\
Orbital distance at periastron $a_{\rm per}$ [AU] & $0.266_{-0.014}^{+0.013}$   \\
Orbital distance at apoastron $a_{\rm apo}$ [AU] & $0.797^{+0.032}_{-0.054}$   \\
Equilibrium temperature at the average distance $T_{\rm eq}$ [K]~$^b$  & $347 \pm 12$ \\
\hline       
\hline
\vspace{-0.9cm}
\footnotetext[1]{\scriptsize in the planet reference frame } \\
\footnotetext[2]{\scriptsize black body equilibrium temperature assuming a null Bond 
albedo and uniform heat redistribution to the night side.} \\
\end{tabular}
\end{minipage}
\label{starplanet_param_table}  
\end{table*}

\subsection{Parameters of the planetary system}
\label{sect_Parameters_planetary_system}

To determine the KOI-3680 system parameters, 
we simultaneously modelled the {\it Kepler} photometry and the SOPHIE RVs in a Bayesian framework.  
We did not include a Rossiter-McLaughlin model because no RVs were obtained during a 
KOI-3680b transit. We employed a differential evolution Markov chain Monte Carlo (DE-MCMC) technique 
(Ter Braak~\cite{terbraak06}, Eastman et al.~\cite{eastman13}) 
 using a Keplerian orbit and the transit model of Mandel \&\ Agol~(\cite{mandel02}).  
Following Bonomo et al.~(\cite{bonomo15}), 
we normalized the transits after accounting for the stellar crowding values provided 
by the \emph{Kepler} team for each quarter, and oversampled the transit model at 1~min to overcome 
the smearing effect due to the long-cadence integration times on the determination of transit parameters 
(Kipping~\cite{kipping10}).
The activity level of the star deduced from the low-amplitude flux variability 
allows us to assume that possible unocculted starspots and/or faculae 
do not significantly affect  the measurement of the transit depth
(Sect.~\ref{sect_kepler_photometry}).
Given the relatively large star--planet distance of $0.33$~AU at inferior 
conjunction, we accounted for a 
light travel time of $\sim 2.7$~min 
between the \emph{Kepler} transit observations 
(which  refer to the planet reference frame) and 
the RV observations (which  refer to the stellar reference frame).

\begin{figure}[h]
\centering
\includegraphics[width=7.1cm, angle=90]{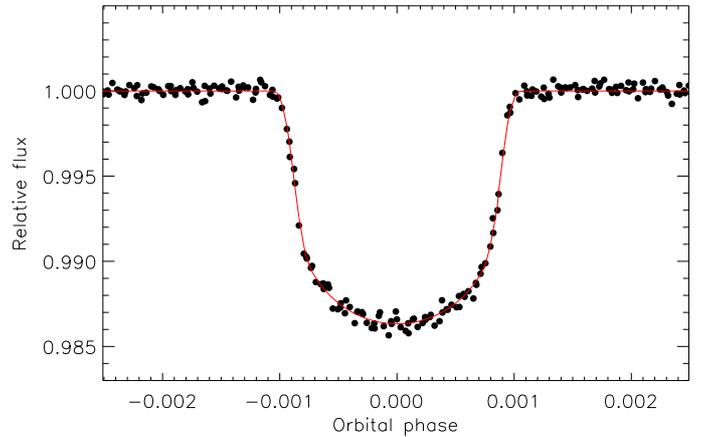}
\caption{Phase-folded transit of KOI-3680b along with the best-fit model (red solid line).}
\label{fig_phase}
\end{figure}

Our global model has twelve free parameters: the mid-transit time $T_{\rm c}$;
the orbital period $P$;
the systemic RV $\gamma$;
the RV semi-amplitude $K$; $\sqrt{e}\cos\omega$ and $\sqrt{e}\sin\omega$,
where $e$ and $\omega$ are the orbital eccentricity and argument of periastron;
the RV uncorrelated jitter term $s_{\rm j}$ (e.g., Gregory~\cite{gregory05});
the transit duration $T_{14}$;
the scaled planetary radius $R_p/R_{\star}$; the orbital inclination $i_p$;
and the two limb-darkening coefficients $q_1$ and $q_2$,
which are related to the coefficients $u_1$ and $u_2$ of the quadratic limb-darkening law
(Claret~\cite{claret04}, Kipping~\cite{kipping13}). 
We used uninformative priors on all the parameters with bounds of [0, 1[ for 
the eccentricity and [0, 1] for the $q_1$ and $q_2$ limb-darkening parameters (Kipping~\cite{kipping13}). 
We ran 24 chains, which is twice the number of free parameters. 
The step orientations and scales for each chain were automatically determined from two of the other chains 
that were randomly selected at each step (Ter Braak~\cite{terbraak06}); 
a proposal step was accepted or rejected 
according to the Metropolis-Hastings algorithm, by using a Gaussian likelihood function.
We adopted the prescriptions given by Eastman et al.~(\cite{eastman13}) 
for the removal of burn-in steps and the convergence check. 
The best-fit models of the radial velocities and transits are displayed in Figs.~\ref{fig_RV} and \ref{fig_phase}. 
The radial velocities fit well with the photometric period and phase, and show no hints of additional signals. 
Any additional RV drift, which could be the signature of a long-period companion in the system, 
should be smaller than~15\,m/s/yr.

The stellar density derived from the transit fitting was used as a proxy for stellar luminosity,  
along with the stellar metallicity and effective temperature
initially derived in Sect.~\ref{sect_stellar_parameters}. They were used 
to determine the stellar parameters, that is, mass, radius, and age, through the Yonsei-Yale 
evolutionary tracks (Demarque et al.~\cite{demarque04}; 
see Bonomo et al.~\cite{bonomo14} for further details). 
Fully consistent stellar parameters were also found with the Dartmouth stellar models 
(Dotter et al.~\cite{dotter08}). 
The final values and $1\sigma$ uncertainties of the stellar, orbital, and planetary parameters 
were computed as the medians and the $[15.87-84.13]$\,\%\ interval of their posterior distributions 
and are listed in Table~\ref{starplanet_param_table};
more specifically, we considered 34.13\,\%\ quantiles above the medians and 34.13\,\%\ 
quantiles below the medians to allow for possibly asymmetric error bars.
The stellar gravity derived here (log\,$g = 4.473_{-0.034}^{+0.027}$ cgs) agrees 
with that obtained from the spectral analysis above 
(log\,$g = 4.49 \pm 0.13$; Sect.~\ref{sect_stellar_parameters}) but is more precise;
this shows the good consistency of those~studies.

These values also 
agree with those from the Gaia DR2 (Gaia Collaboration~\cite{gaia18}). 
More specifically, 
our derived stellar radius $R_\star=0.96_{-0.06}^{+0.05}~\Rsun$ is compatible within $1\sigma$
of that estimated from the Gaia data, that is, $R_\star=1.01 \pm 0.05~\Rsun$
(Berger et al.~\cite{berger18}). However, we report the latter estimate 
for comparison and do not consider it 
necessarily more reliable than our value of $R_\star$, because it was derived 
from a bolometric correction that in turn relies on stellar models (Berger et al.~\cite{berger18}). 

From our stellar parameters, the radius ratio $R_{\rm p}/R_{*}$, and the RV semi-amplitude, we find 
that KOI-3680b has a radius $R_{\rm p}=0.99^{+0.06}_{-0.07}\,\rm R_{Jup}$, 
a mass $M_{\rm p} = 1.93_{-0.21}^{+0.19}\,\rm M_{Jup}$, 
and thus a density $\rho_{\rm p}=2.46^{+0.42}_{-0.36}$\,g\,cm$^{-3}$. 
As mentioned above, the planet revolves around its host star on a considerably eccentric~orbit, 
$e=0.496 \pm 0.031$ (see Fig.~\ref{fig_RV}). The system age is~unconstrained.
Finally, the stellar and planetary parameters provided by the DR25 \kepler\ catalogue 
(Thompson et al.~\cite{thompson18}) also agree with~ours.

\subsection{Transit timing variations}

The 141-day period is known with an uncertainty of only $\pm7.4$\,seconds. The tremendous 
accuracy offers an opportunity to search for any possible TTV. Such a TTV could be 
caused by additional companions and reveal a multiple planetary~system. 

\begin{figure}[h]
\centering
\includegraphics[width=6.6cm, angle=90]{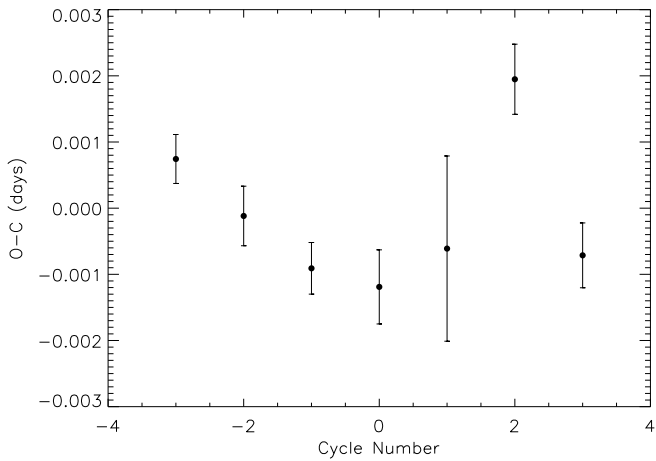}
\caption{Transit timing variations of KOI-3680b with respect to the linear 
ephemeris reported in Table~\ref{starplanet_param_table}.}
\label{fig_TTV}
\end{figure}

In the same DE-MCMC Bayesian framework, we computed the 
individual transit times which are listed in Table~\ref{table_TTV}.
The transit epoch with the highest uncertainty refers to the fifth partial 
transit (Sect.~\ref{sect_kepler_photometry}).
The derived TTVs and their error bars are shown in Fig.~\ref{fig_TTV}.
They are similar to those determined independently by 
Holczer et al.~(\cite{holczer16}) and they
present some marginal variations that would require more investigations 
and possible transit re-observations, which go beyond the scope of the present~work. 
With the present data and analyses, we only detect and characterize the new planet KOI-3680b and
could not report any significant detection of additional companions in the~system.

As a sanity check, we performed an additional combined analysis as in 
Sect.~\ref{sect_Parameters_planetary_system} after correcting for the
marginal mid-transit variations. This shown no significant differences in the 
resulting system~parameters.

\begin{table}[h]
\caption{Times of KOI-3680b mid-transits.}            
\begin{center}
\renewcommand{\footnoterule}{}                          
\begin{tabular}{l l}      
\hline \hline
Time & Uncertainty  \\
$[\rm BJD_{TDB}-2\,450\,000]$ & [days] \\
\hline
5512.93007      &       0.00037 \\
5654.17088      &       0.00045 \\
5795.41176      &       0.00039 \\
5936.65315      &       0.00056 \\
6077.8953       &      0.0014     \\
6219.13963      &       0.00053 \\
6360.37864      &       0.00049 \\
\hline \hline
\label{table_TTV}
\end{tabular}
\end{center}
\end{table}

\section{Discussion}
\label{sect_discussion}

Being a giant planet with a five-month orbital period, KOI-3680b occupies 
a position in the parameters space 
where only a few planets are known today, in particular the transiting ones. 
The mass distribution in the total population of known exoplanets is bimodal with a 
gap around $M_{\rm p} \sin i_{\rm p} = 0.15$~\MJ. 
The distinction between those two populations of low-mass and giant planets is well known from 
RV surveys (e.g. Mayor et al.~\cite{mayor11}). 
The mass valley has a corresponding 
radius valley seen in the population of transiting planets.
These valleys are likely to be due to 
the ways giant planets grow depending on the mass of their core 
(e.g. Mordasini et al.~\cite{mordasini12}). It could be also partially 
due to evaporation processes and interactions with the host stars 
in the case of close-in planets (e.g. Owen~\&\ Wu~\cite{owen17}, 
Fulton et al.~\cite{fulton17}, 
Van~Eylen et al.~\cite{vaneylen18}).
The mass and radius of KOI-3680b  put it in the giant planet~population.

On the other hand, known giant planets  show a bimodal distribution in their orbital periods, 
with close-in hot Jupiters at short periods and temperate giants at longer periods. The intermediate 
domain (periods between 10 and 100 days) shows a dearth of planets.
That period valley was first identified by RV surveys (e.g. Udry et al.~\cite{udry03})
but was difficult to detect in the case of transiting planets (Santerne et al.~\cite{santerne16}) as
the latter method could be polluted by false positives and 
highly favours the detection of close-in planets.
Indeed, only a few transiting, giant planets are known on orbital periods longer than a few tens of days.
KOI-3680b is one of these rare~cases.

\begin{figure}[h]
\centering
\includegraphics[width=8.8cm, angle=0]{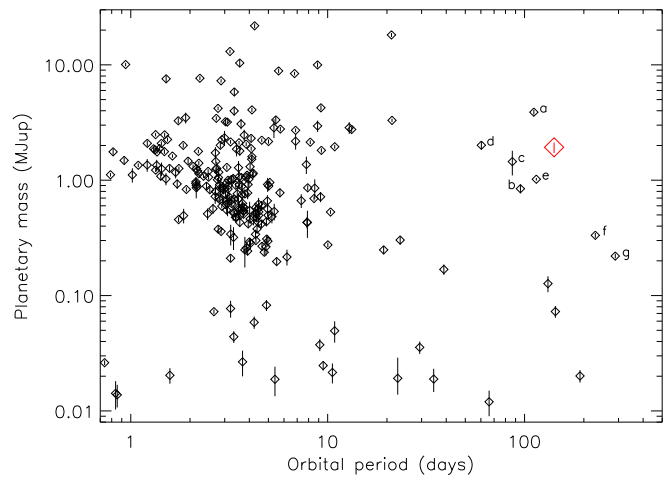}
\caption{Masses of transiting exoplanets as a function of their orbital periods.
The figure only shows planets with radius and mass  measured at better than $\pm30$\,\%\ each.
The new planet KOI-3680b is plotted in red.
Other planets discussed in the text are identified in the plot by letters 
$a$ (HD\,80606b),
$b$ (CoRoT-9b),
$c$ (KOI-1257b), 
$d$ (Kepler-30c),
$e$ (Kepler-87b),
$f$ (Kepler-16b), and 
$g$ (Kepler-34b).}
\label{fig_perVSmass}
\end{figure}

\begin{figure}[h]
\centering
\includegraphics[width=8.8cm, angle=0]{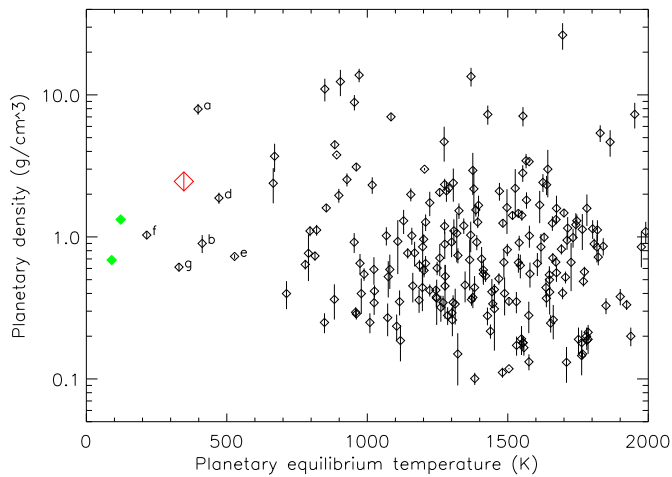}
\caption{Bulk density of giant planets as a function of their equilibrium temperature.
Only known planets with a mass larger than 0.15\,\MJ\ and a density measured at better than $\pm50$\,\%\ 
are plotted here.
The new planet KOI-3680b is plotted in red, and Saturn and Jupiter in filled green.
Planets are labelled as in Fig. 7.}
\label{fig_tequVSdensity}
\end{figure}

Whereas several known transiting exoplanets have orbital periods longer than that of KOI-3680b 
(Rowe et al.~\cite{rowe14}, Morton et al.~\cite{morton16}, Giles et al.~\cite{giles18}), 
only a few of them are well characterized, in particular with a mass measurement.
Measured masses and orbital periods of transiting planets are plotted in 
Fig.~\ref{fig_perVSmass}\footnote{Figures~\ref{fig_perVSmass} and \ref{fig_tequVSdensity}
are based on the Exoplanet Orbit Database at exoplanets.org (Han et al.~\cite{han14}).}, where 
giants are on the upper part. Plotted in red, KOI-3680b has parameters similar to 
five other giant planets: HD\,80606b ($P=111.4$\,d), CoRoT-9b ($P=95.3$\,d), and KOI-1257b
($P=86.6$\,d) with masses also measured from radial velocities (H\'ebrard et al.~\cite{hebrard10},
Bonomo et al.~\cite{bonomo17a}, Santerne et al.~\cite{santerne14}), 
and Kepler-30c ($P=60.3$\,d) and Kepler-87b ($P=114.7$\,d)
with masses measured through TTVs 
(Fabrycky et al.~\cite{fabrycky12}, Ofir et al.~\cite{ofir14}).
In that six-planet group, KOI-3680b has the longest period.
Two giant planets with slightly lower mass have longer periods in Fig.~\ref{fig_perVSmass}: 
Kepler-16b ($P=228.8$\,d) and Kepler-34b ($P=288.8$\,d) which 
are circumbinary planets with TTV-measured masses
(Doyle et al.~\cite{doyle11}, Welsh et al.~\cite{welsh12}).
Therefore, to date, KOI-3680b is the known transiting giant planet with the longest period 
characterized around a single star.

Larger distances to the host stars imply lower planetary temperatures. In the case of KOI-3680b, 
the black-body equilibrium temperature at the average distance is $T_{\rm eq} = 347 \pm 12$\,K
(310\,K and 530\,K at apastron and periastron, respectively), 
assuming a null Bond albedo and uniform heat redistribution to the night side 
(e.g. Alonso et al.~\cite{alonso09}). Figure~\ref{fig_tequVSdensity} shows equilibrium 
temperatures $T_{\rm eq}$ of known giant planets and their corresponding bulk densities.
The plotted $T_{\rm eq}$ should be taken with caution as the actual $T_{\rm eq}$ values 
would depend on albedos and heat redistributions which vary from one planet to another, 
as well as on star--planet distances which vary for eccentric orbits. Nevertheless, the $T_{\rm eq}$ 
values plotted in Fig.~\ref{fig_tequVSdensity} provide useful orders of magnitude
and indicate that KOI-3680b is one of the rare temperate planets to be well characterized
with measured mass and density.
The distribution suggests 
a decreasing lower envelope which shows that the less-dense 
giant planets have higher temperatures.
This trend should be taken with caution as there are only a few points below 
$T_{\rm eq} = 800$\,K.~Still,
Jupiter and Saturn are both 
located near that lower envelope. 
If real, this trend would agree
with the 
tendency of hotter planets to have inflated radii (e.g. Fortney et al.~\cite{fortney07}, 
Enoch et al.~\cite{enoch12}).

In Fig.~\ref{fig_tequVSdensity}, only two exoplanets have $T_{\rm eq}$ lower than that of 
KOI-3680b: they are the circumbinary planets Kepler-16b and Kepler-34b.
Three others have slightly higher $T_{\rm eq}$ and lower densities: CoRoT-9b, Kepler-30c, and 
Kepler87b. The planet HD\,80606b has a slightly larger $T_{\rm eq}$ than that of KOI-3680b but a 
density three times greater. KOI-3680b is therefore among 
the colder giant planets characterized around a single star.

\begin{figure}[h]
\centering
\includegraphics[width=6.6cm, angle=90]{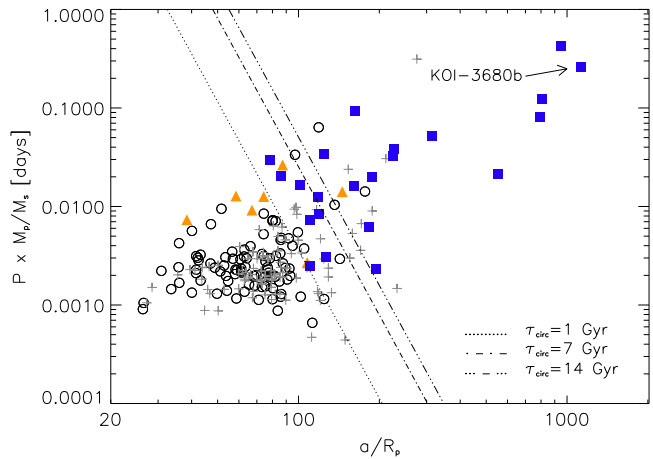}
\caption{`Modified' tidal diagram for 232 giant planets with uniformly derived eccentricities 
(Bonomo et al.~\cite{bonomo17b}). 
Black empty circles show the position of giant planets with well-determined circular orbits, 
i.e. with eccentricities consistent with zero and $1\sigma$ uncertainty $\sigma_{e}<0.05$; 
grey crosses indicate planets with undetermined eccentricities 
which are usually consistent with zero but have large uncertainties $\sigma_{e}>0.05$; 
orange triangles show small but significant eccentricities, i.e. $e < 0.1$; and blue squares $e \ge 0.1$. 
The position of KOI-3680 is indicated in the upper-right corner. 
The dotted, dash-dotted, and dash-three-dotted lines display the 1, 7, and 14-Gyr circularization timescales 
for a planetary modified tidal quality factor $Q'_{\rm p}=10^6$ and $e=0$, respectively.}
\label{fig_tidal_diagram}
\end{figure}

Figure~\ref{fig_tidal_diagram} shows the position of KOI-3680b
in the `modified' tidal diagram of transiting giant planets 
whose orbital eccentricities were uniformly determined by Bonomo et al.~(\cite{bonomo17b}).
In this diagram the $(P \times M_{\rm p}/M_{\rm s})$ quantity is 
plotted as a function of $(a/R_{\rm p})$; since the circularization time scales as 
$\tau_{\rm circ} \propto P \times M_{\rm p}/M_{\rm s} \times (a/R_{\rm p})^5$, eccentric orbits 
(blue squares for $e~\ge~1$ in Fig.~\ref{fig_tidal_diagram}) are mostly expected at the right or 
upper-right part of the diagram, being almost unaffected by tidal dissipation/circularization. 
Indeed, they all lie beyond the 1~Gyr circularization isochrone, and two-thirds of them beyond 
the 14~Gyr one (dash-three-dotted line in Fig.~\ref{fig_tidal_diagram}). 
KOI-3680b is currently the transiting giant planet with the largest $a/R_{\rm p}$ and 
mass measured through radial velocities, and is located at the upper-right corner of this diagram.

The high eccentricity of KOI-3680b does likely not originate from
interactions with the protoplanetary disc, which tend to damp high 
eccentricities (e.g. Kley \&\ Nelson~\cite{kley12}, Bitsch et al.~\cite{bitsch13}). 
Instead, it must be the outcome of evolution processes occurring after the disc dissipation, 
such as planet--planet gravitational scattering (e.g. Chatterjee et al.~\cite{chatterjee08}), 
Kozai-Lidov perturbations caused by a distant stellar or planetary companion on a highly inclined
orbit (e.g. Fabrycky \&\ Tremaine~\cite{fabrycky07}), 
and/or secular chaos in a multi-planet system (e.g. Wu \&\ Lithwick~\cite{wu11}). 
Carrying on the RV follow-up of this system might unveil additional outer companions 
and clarify which is the most likely mechanism that gave rise to the high orbital eccentricity,  
as is the case for CoRoT-20b (Rey et al.~\cite{rey18}).

In terms of interior structure and evolution, KOI-3680b appears normal, that is, given the large 
uncertainty on its age, evolution models (Guillot~\cite{guillot05}, Guillot et al.~\cite{guillot06}) 
are able to match the observed radius both when assuming a solar composition and no core 
or a massive central core up to $140\,\rm M_\oplus$. The contraction of the planet is still relatively 
significant (compared to hot Jupiters for which cooling is slowed by their irradiation): 
the planet contracted by about $0.06$\,\Rjup\ between the ages of 1 and 10 Gyr. 
This implies that a more precise determination of the stellar age (and generally of 
stellar parameters) would certainly yield useful constraints on the composition of this planet.  

The new transiting planet KOI-3680b
could allow characterizations in the mostly unexplored domain of temperate planets.
Whereas most of the atmospheric studies are made in close-in planets, they could be feasible 
here in a more temperate domain, closer to that of Jupiter or Saturn. 
They would require extremely large telescopes however, such as for example the Thirty Meter Telescope, 
due to the faintness of the host star.  Still, due to the lower 
temperature, the scale height of the atmosphere is expected to be two to four times smaller 
than for close-in, giant planets, resulting in a  shallower  absorption signal. The emitted light 
detectable through occultations would also be smaller than that of hotter planets.
In terms of obliquity measurement, the amplitude of the Rossiter-McLaughlin anomaly 
is expected to be of the order of 35~m/s. Such accuracy is barely reachable with SOPHIE 
due to the magnitude of the star, but this is feasible with a stable spectrograph mounted on 
a larger, more sensitive telescope such as HARPS-N at TNG or HIRES at Keck.
The long duration of the transits (6.7~hours) 
allows for more RV measurements to be obtained during a transit,
but make it harder to find a transit fully observable during a given night.
Combined observations of a given transit with different telescopes shifted in longitude 
or observations  of different transits could help here~(e.g.~H\'ebrard~et~al.~\cite{hebrard10}).

Finally, with its long orbital period in comparison to most  known transiting giant planets, 
KOI-3680b has a large Hill sphere of about 3.5~million kilometers, where the gravity 
of the planet dominates the gravity of the star. 
That extended sphere of gravitational influence makes the presence of satellites 
and rings around KOI-3680b more likely.
Indeed, satellites must be included well within the Hill sphere to have stable orbits, and stable
prograde orbits are typically within about 0.4~times the radius of the Hill sphere 
(e.g. Hinse et al.~\cite{hinse10}).   
In addition, giant planets are more likely to host satellites and rings, as in the solar system. 
We note however that if the high eccentricity of KOI-3680b is due to dynamical instabilities, 
this may be detrimental for possible moons 
(e.g. Bonomo et al.~\cite{bonomo17a}, Lecavelier des \'Etangs et al.~\cite{lecavelier17}).
Nevertheless, KOI-3680b is today one of the rare known systems  favourable for the search of such 
structures, whereas most transiting planets have short periods and extremely tight Hill spheres 
of a  few planetary radii.
This probably explains the lack of satellites and rings detection outside the solar system.
Today, the only known cases are the possible exomoon Kepler-1625b\,I 
(Teachey \&\ Kipping~\cite{teachey18})
and the suspected ring system in J1407 (Mamajek et al.~\cite{mamajek12}).

\section{Conclusions}
\label{sect_conclusion}

Following identification by the \kepler\ team as a promising candidate, we have 
established the planetary nature of the KOI-3680.01 transits and characterized 
this exoplanetary system. Its orbital period of 141~days, located between those of Mercury 
and Venus, makes KOI-3680b the known transiting, giant planet with the longest period characterized 
today around a single star. 
It offers opportunities to extend follow-up studies of exoplanets farther from 
their stars, and to compare close-in and more temperate planets.
After the~announcement of the present results, the \kepler\ Team gave the system 
KOI-3680 the name \object{Kepler-1657} (Table~\ref{startable_KOI}).

With its four-year continuous observations, the \kepler\ mission was one of the rare experiments 
particularly favourable to the detection of long-period, transiting planets. Other programs 
including CoRoT, K2, Cheops, and ground-based surveys are less sensitive to such
a range of periods due to the shorter durations of their continuous observations. 
Mainly PLATO (Rauer et al.\cite{rauer14})
and surveys secured from Antartica (e.g. Crouzet et al.~\cite{crouzet18}) 
could be able in the future to detect 
numerous transiting planets on longer periods.
Whereas it monitors most of the sky with durations of  27~days only, the recently launched 
TESS satellite acquires photometry on longer time spans for a few parts of the~sky, and therefore 
might also detect some long-period, transiting planets (e.g. Sullivan et al.~\cite{sullivan15}, 
Villanueva et al.~\cite{villanueva18}) which could eventually be characterized.

\begin{acknowledgements}
This publication is based on observations collected with the NASA \kepler\ mission and 
the SOPHIE spectrograph on the 1.93-m telescope at Observatoire de Haute-Provence (CNRS), 
France.
We thank all the OHP staff of for their support.
This work was  supported by the ``Programme National de Plan\'etologie'' (PNP) of CNRS/INSU, 
the Swiss National Science Foundation, 
the French National Research Agency (ANR-12-BS05-0012),
FEDER - Fundo Europeu de Desenvolvimento Regional funds through the COMPETE 2020 - 
Programa Operacional Competitividade e Internacionaliza\c{c}\~ao (POCI), and by 
the Portuguese funds through FCT - Funda\c{c}\~ao para a Ci\^encia e a Tecnologia in the framework 
of the projects POCI-01-0145-FEDER-028953 and 
POCI-01-0145-FEDER-032113. 
N.\,C.\,S. further acknowledges the support from FCT through national funds and by FEDER through 
COMPETE2020 by these grants UID/FIS/04434/2013 \&\ POCI-01-0145-FEDER-007672.
A.S.B. acknowledges funding
from the European Union Seventh Framework Programme (FP7/2007-2013)
under Grant Agreement No. 313014 (ETAEARTH). 
S.C.C.B. acknowledges support from FCT through Investigador FCT 
contract IF/01312/2014/CP1215/CT0004.
\end{acknowledgements}

\end{document}